# Revised masses of dust and gas of SLUGS FIR bright galaxies based on a recent CO survey

Ernest Seaquist,[1] Lihong Yao,[1] Loretta Dunne[2] and Heather Cameron[1]

[1]*Department of Astronomy & Astrophysics, University of Toronto, Toronto, Ontario, Canada M5S 3H8*
[2]*Department of Physics and Astronomy, University of Wales Cardiff, PO Box 913, Cardiff CF2 3YB*

Accepted __________

**ABSTRACT**
Recent CO measurements of an essentially complete sub-sample of galaxies from the SCUBA Local Universe Survey (SLUGS) are used to examine their implications for dust and gas masses in this sample. Estimates of dust masses are affected by a contribution to the SCUBA brightness measurements by CO(3-2) emission, and molecular gas masses by the use of a modified value of the CO-to-$H_2$ conversion factor X. The average dust mass is reduced by 25-38 per cent, which has no bearing on earlier conclusions concerning the shapes of the dust mass luminosity function derived from the SLUGS. The value of X found from the CO survey, when applied together with the reduction in dust masses, leads to lower estimates for the mean gas-to-dust mass ratios, where the gas includes both $H_2$ and HI. For the CO sample, the mean global ratio is reduced from approximately 430 to about 320-360, but is further reduced to values near 50 when applied to the nuclear regions relevant to the CO observations. We discuss these results and suggest that the differences between the nuclear and outer regions may simply reflect differences in metallicity orthe existence of considerable amounts of unobserved cold dust in the outer regions of these galaxies.

Key words: dust, gas, masses - galaxies: ISM - galaxies: starburst - infrared: galaxies

## 1 INTRODUCTION

The measurement of reliable masses for the dust and gas in galaxies requires a reliable method of interpreting IR/sub-mm continuum emission for the dust, and CO molecular line emission for the $H_2$ molecular gas. In this paper we re-examine the gas-to-dust mass ratios of a representative sub-sample of galaxies contained in the 850μm SCUBA Local Universe Survey (SLUGS) by Dunne et al. (2000) in the light of a CO survey of about half this sample presented by Yao et al. (2003). The SLUGS survey, conducted with the James Clerk Maxwell Telescope (JCMT) contains 104 galaxies selected according to their IRAS 60μm flux, and constitutes a flux limited sample with far-IR luminosities in the range $10^{10}$ $L_{\odot} < L_{FIR} < 10^{12}$ $L_{\odot}$. The CO survey, also conducted with the JCMT, contains about 60 galaxies over a restricted range in R.A., and is essentially complete to the same IRAS flux. Both CO(1-0) and CO(3-2) measurements in this study have the same angular resolution (~15") as the SLUGS survey, but the CO survey includes only single pointings at the nominal galaxy position, whereas the SLUGS survey contains complete images at 850μm. The discussion in this paper focuses on the effects of contamination by CO(3-2) and the potential for contamination by CO(6-5) of SCUBA fluxes at respectively 850μm and 450μm, and on inferences from adopting a lower than conventional value for the CO-to-$H_2$ conversion factor X found by Yao et al. (2003) in their analysis of the CO survey. This factor is commonly used to scale the brightness of CO(1-0) to $H_2$ column density.

### 1.1 Contamination of sub-mm continuum by molecular line emission

The interpretation of sub-mm continuum observations of galactic and extragalactic sources requires that allowance be made for contamination by line emission within the bolometer bandpass. In particular, several authors have pointed out that observations made with SCUBA on the JCMT are likely to include significant line emission. For example, Johnstone & Bally (1999) find that the contribution by all line emission at the brightest peak in the northern part of OMC-1 is 28 per cent at 850μm and 24 per cent at 450μm, mainly due to SO and $SO_2$. In many sources, however, the primary source of contaminating emission is CO(3-2) at 850μm and potentially CO(6-5) at 450μm. The centre of the 850 μm bandpass filter is about 347 GHz corresponding closely to the rest frequency of the CO(3-2) transition at 345.796 GHz, and the corresponding figures for the current 450 μm filter and the CO(6-5) transition are respectively 664 GHz and 691.473 GHz. Davis et al. (2000) estimate that contamination of SCUBA emission by CO(3-2) ranges between 10-50 per cent in the outflow of the protostar V380Ori-NE. Tothill (1999) deduced that line emission produces 10-50 per cent of the continuum emission seen in the Lagoon Nebula. A number of estimates have been made for starburst galaxies as well. For NGC 7469, Papadopoulos & Allen (1998) estimate that the contribution to the 850μm band by CO(3-2) is 40 per cent, but that the contribution at 450μm by CO (6-5) is negligible. For the prototypical starburst galaxy M82, Seaquist & Clarke (2001) estimated that CO contributes up to 47 per cent of the 850μm continuum, and Zhu, Seaquist & Kuno (2003) estimate that CO contributes about 46 per cent overall to the 850μm continuum in the interacting system NGC 4038/4039. Because the corrected 850μm emission corresponds to a higher temperature for cool dust, corrections to the dust mass can be a factor of several when the contamination is near 50 per cent (eg. Zhu et al. 2003).

However, these isolated examples do not represent a systematic investigation of the contamination by CO for a representative sample of galaxies. Nor do they address the issues of the uncertainties in the measured contamination. The work described here investigates the consequences of the fractional contamination by CO in a systematic way, and assesses the consequences.

### 1.2 The effects of the CO-to-$H_2$ conversion factor X

It is conventional to derive masses of $H_2$ gas in our own Galaxy, and in other galaxies, by applying a conversion factor X to derive the $H_2$ column density from the CO(1-0) brightness integrated over velocity. However, it is strongly suspected that this conversion factor is not uniform, and may depend upon both the metallicity and physical conditions in the galaxy. A more complete discussion of this issue is contained in Yao et al. (2003) and references therein. The empirical value normally adopted for GMC’s in our own Galaxy is X = $2.8\times10^{20}$ $cm^{-2}$ $(K\ kms^{-1})^{-1}$ (hereafter referred to as the conventional value). In keeping with a prevailing view that the value for star-forming galaxies is systematically lower than this, Yao et al. (2003) used their CO measurements together with LVG modeling methods (Goldreich & Kwan 1974) to derive a value for the SLUGS sample in the range $1.3\times10^{19}$ - $6.6\times10^{19}$ $cm^{-2}$ $(K\ kms^{-1})^{-1}$, ie a factor of 4 - 20 lower than the conventional value. In this paper, we also discuss the implications of the lower values for the estimates of gas-to-dust mass ratios in the SLUGS sample based on the revised X factor and the analysis of the contamination of the 850μm continuum emission by CO(3-2).

## 2 DATA AND ANALYSIS

### 2.1 Contribution by CO(3-2) to 850μm emission

The contribution to the main beam continuum surface brightness S(V) by line emission, expressed in flux units per beam area may be written,

$$S(V) = \frac{2k\ {}_{L}^{2}}{c^2}\ {}_{b}\ \frac{g(V)}{\int g(V)\,dV}\ I_{CO} \qquad (1)$$

where ${}_{L}$ is the line frequency,

${}_{b}$ is the main beam solid angle,

g(V) is the bandpass profile of the filter, expressed as a function of recession velocity V = cz,

$I_{CO} = \int T_{mb}\, dV$ is the velocity integrated main beam brightness temperature of the CO line.

There have been two sets of SCUBA filters on the JCMT since November 1999. The bandpass profiles for all filters were measured in-situ in December 2000 by David Naylor (2002). The older set, relevant to the SLUGS data, is of primary concern to us here, although the new 850μm profile does not differ significantly from the old one. We discuss the contamination in the new and old 450μm filters in section 2.2. The profile of the old 850μm filter was provided to us by David Naylor and is shown in Figure 1. The profile has been smoothed by a boxcar window (ie a running mean) of width 300 km $s^{-1}$, comparable to mean Doppler width of the CO lines in the SLUGS sample. The smooth curve through the data represents a double gaussian fit, by least squares, leading to a functional form for $S(V) = F(V)\, I_{32}$, expressed in mJy $beam^{-1}$. Here F(V) is given by

$$F(V) = 1.101 \exp \left\{ -0.5 \left( \frac{V + 5122}{6057} \right)^2 \right\} + 0.612 \exp \left\{ -0.5 \left( \frac{V - 9201}{3420} \right)^2 \right\} \qquad (2)$$

where V is in km $s^{-1}$. $I_{32}$ is the intensity of the CO(3-2) line, expressed in K km $s^{-1}$, and $_{b}$ = $5.6 \times 10^{-9}$ sr corresponding to the FWHM beamsize of ~15" of the JCMT at 850μm.

The excursions about the best fit curve, ie the bandpass ripple, are produced by reflections in the filter cavity, and it is not certain whether these are stable in amplitude or position. The rms deviation from the curve is 24 per cent in the velocity range $0 < V < 10{,}000$ km $s^{-1}$, and this value was assigned as the corresponding uncertainty in converting line emission to S(V). The CO(3-2) brightness values are contained in Table 1 of Yao et al. (2003) and converted to S(V) by applying equation (2). Each CO measurement is based on a single pointing corresponding to the pixel of the SCUBA image at the centre of the galaxy. The corresponding 850μm brightness corresponds to the central pixel of the SCUBA image. The uncertainty in the relative pointing is about 3.6" which may be compared with the FWHM beamsize of ~15". The uncertainty due to noise in the SCUBA brightness values was set equal to the rms in blank regions of the map combined quadratically with an assumed uncertainty in the gain of the telescope of 10 per cent. Since the region sampled by the beam is fixed by the angular resolution of ~15", the projected linear size sampled ranges from 2 - 20 kpc over the survey sample.

Figure 2 shows a plot of S(V) vs. SCUBA brightness for the CO sub-sample. Arp 220 was not included in the plot because of its very high value SCUBA brightness (667 mJy). The corresponding value of S(V) for this galaxy is 42 mJy. The data are segregated according to $V < 10{,}000$ km $s^{-1}$ (filled circles) and $V > 10{,}000$ km $s^{-1}$ (open circles). This velocity was used to mark the point beyond which S(V) declines rapidly with increasing velocity. The line through the data has slope $0.26 \pm 0.01$ representing a linear regression with zero intercept to data with $V < 10{,}000$ km $s^{-1}$. The fit also excluded Arp 220 and the three points in Figure 1 with large scatter at high SCUBA brightness, since otherwise they would receive unduly large weight in the fit. The fit was done using the bisector method (e.g. Akritas & Bershady 1996) allowing for error in both coordinates. The slope represents one estimate for the mean fractional contribution of CO line emission.

The data were also analyzed in a different way by forming the ratio of S(V) to SCUBA brightness directly, restricting the sample to measurements with S/N of at least 3 . This restricted the sample to 40 objects, primarily due to the poor S/N ratio on some of the SCUBA single pixel measurements. However, this analysis includes Arp 220 and the other points left out in the regression analysis. The distribution of the ratios is plotted in Figure 3. The mean value for V < 10,000 km $s^{-1}$ (36 points) is 0.24 with rms of 0.09 about this mean. To examine whether the observed variation in the ratio is real or dominated by measurement errors, we show superposed on the histogram a curve representing the expected distribution of the ratio using a true value of 0.24 with mean standard errors of 32 and 25 per cent for S(V) and SCUBA brightness respectively. These errors are assumed each to possess a gaussian random distribution, and the curve takes account of the asymmetry introduced by taking the ratio of two numbers when the errors are non-negligible fractions of the true value. The similarity between the curve and the histogram suggests that the observed distribution is dominated by the presence of measurement errors, so that it is unlikely that the range found in our measurements represents a true range among the galaxies.

Figure 4 shows a plot of the ratio of CO(3-2) line to SCUBA continuum flux vs. FIR luminosity from Dunne et al. (2000). Note that in this case the line fluxes are not folded through the filter profile to obtain S(V), so that the recession velocity has no effect on the ratio. Since FIR luminosity and distance are very strongly correlated in these data, the figure tests for a possible dependence of contamination by CO on either FIR luminosity or projected beam size at the galaxy. The scatter in these data is large, as noted in Figure 3, and we find no significant correlation. The linear correlation coefficient is -0.14 at the significance level of 38%. Thus errors of measurement mask any relationship between CO contamination and fraction of the galaxy covered by the beam, which varies from about 15 to 100 per cent.

**2.2 Contribution by CO(6-5) to 450µm emission**

SCUBA data taken with the 450µm filter are also subject to potential contamination by CO(6-5). However, the old filter in use during the SLUGS survey is centered at about 750 GHz, and is sufficiently narrow (FWHM = 38 GHz), that line emission occurs well below the half power point of the filter response. The new wider band filter is centered at about 664 GHz with FWHM bandwidth 63 GHz, so line contamination is more likely in this case. Here the contamination ratio was estimated by extrapolating the CO brightness values from intensities at the lower transitions CO(1-0) and CO(3-2). The mean ratio of the integrated intensities CO(3-2)/CO(1-0) is 0.66 (Yao et al. 2003), and Figure 13 of Yao et al. (2003) shows that the corresponding kinetic temperature (assumed equal to the dust temperature) and $H_2$ density are 35K and $2\times10^{3}$ $cm^{-3}$ respectively. Using an LVG code with these parameters and $X_{CO}/dv/dr = 10^{-5}$ km $s^{-1}$ pc, we find an average expected line ratio CO(6-5)/CO(3-2) = 0.032. Using an equivalent filter bandwidth for the new 450µm filter of 67 GHz, and a dust emissivity index =2.0 to scale the continuum intensities between 850 and 450µm, we find an average fractional line contamination ratio less than one per cent of the amount observed at 850µm for a typical galaxy in the SLUGS sample. Thus it is safe to assume that in general no corrections are needed for CO at 450µm for the SLUGS sample.

## 3 DISCUSSION

### 3.1 Effect on dust mass distribution

In this section we examine the impact of the contamination of SCUBA 850μm fluxes by CO(3-2) emission on the dust mass function for the SLUGS galaxies measured by Dunne et al. (2000) and Dunne & Eales (DE01, 2001). In the next section we investigate the more difficult issue of absolute dust and gas masses and the gas-to-dust ratio. As noted by Dunne et al. (2000), there are large systematic uncertainties in determining absolute dust masses from sub-mm measurements, owing mainly to the uncertainty in the absolute value of the dust emissivity at these wavelengths. However, as they also note, the shape of the dust mass distribution concerns only the relative dust masses, and this should not be affected by the uncertainties in the absolute dust masses. This distribution was modeled by Dunne et al. (2000) using the Schechter function, which is commonly used to describe the luminosity function for galaxies (Press and Schechter 1974; Schechter 1975). The isothermal (ie single temperature model) dust masses for the SLUGS galaxies were estimated by Dunne et al. (2000) from their 850μm fluxes according to the equation

$$M_d = \frac{S_{850}\ D^2}{\kappa_{850}\ B(850\mu m,\ T_d)} \qquad (3)$$

where D is the distance given in equation (2) of Dunne et al. (2000), $_{850}$ is the dust mass opacity coefficient at 850μm and B(850μm, $T_d$) is the Planck function for a dust temperature $T_d$. The benefits of using sub-mm data to measure the dust mass are that the dust is optically thin and that the dust opacity is dependent only on the grain volume per unit mass, and not on the distribution of grain sizes. The latter property is true for any wavelength greatly exceeding the grain diameter (e.g. Hildebrand 1983). For $T_d \geq$ 30K temperatures, the Rayleigh-Jeans law is applicable, and the dust mass is approximately proportional to $S_{850}/T_d$. It is more temperature sensitive at lower $T_d$, where at 850μm the Rayleigh-Jeans approximation fails. In Dunne et al. (2000), dust masses were estimated using $_{850} = 0.077\ m^2\ kg^{-1}$, and a temperature derived from fitting an isothermal dust SED to the IRAS and 850μm data points.

For the 40 SLUGS galaxies with central brightness S/N > 3 , a correction was made to the 850μm flux for the CO(3-2) line contamination. Since the uncertainty in the CO(3-2)/850μm flux ratio is very large for any individual galaxy, we have simply corrected all galaxies using a mean ratio of 0.25 (ie the average between those found from Figures 2 and 3). Each galaxy SED was then re-fitted to find $T_d$ and the emissivity index . We found that the index fitted to the SED was higher (as the 850μm flux was lower) and the temperature was *slightly* cooler (by 1-2 K). Since the difference in temperature is small, the fractional change in dust mass at temperatures of 30 - 50 K is equivalent to that in the 850μm flux, ie an average value of ~ 25 per cent.

In Dunne & Eales (DE01, 2001), a more realistic two-temperature SED was fitted to a limited number of SLUGS galaxies which also possess ISO and/or 450μm data. It was found that all of these galaxies required a second, colder dust component at ~ 20K. In addition, a fixed value for =2.0 was strongly suggested by the uniformity of the ratios of 850μm/450μm fluxes. We used the objects from this work to determine how dust masses using two component models are affected by corrections to the 850μm fluxes. In these cases, we found that when fixing =2.0, the 25 per cent downward correction in 850μm flux produced an increase in cold dust temperature by a few K. However the fractional decrease in dust mass is larger than 25 per cent because the Rayleigh-Jeans limit is not as readily applicable at $T_d$ = 20K. The revised dust masses for these galaxies average about 38 per cent lower than in DE01, with a range of 22-52 per cent. With either model, the effect on the dust mass functions presented in Dunne et al. (2000) and DE01 is quite straight-forward, however, since no correlation was found between the fractional size of mass correction required and the dust mass of the object. The dust mass functions are thus still well fit by Schechter functions as in Dunne et al. (2000). The effect of the corrections is simply a decrease in M*, the scaling parameter analogous to Schechter parameter L*, which determines the turnover in the dust mass distribution. The parameter M* decreases by either 25 per cent in the case of the isothermal masses presented in Dunne et al. (2000), or by 38 per cent for the cold dust mass function presented in DE01.

One caveat to note is that the CO(3-2) fluxes and corresponding 850μm fluxes refer to only the central ~15" region of each galaxy. The fractional contamination of the 850μm flux was derived for this region and then extrapolated over the rest of the galaxy using fully sampled SCUBA maps from the SLUGS database. If the CO(3-2)/850μm ratio varies with galaxy radius, then obviously our corrections here will be incorrect. Mapping of a large sample of objects in the CO(3-2) line would be needed to fully address this question. However, we do note that there appears to be at most a weak systematic trend of the CO(3-2)/850μm ratio with projected beam size evident in Figure 4.

### 3.2 Gas-to-dust mass ratios

The gas-to-dust mass ratio (g/d) is an important factor in understanding galaxy evolution, since it reflects both the metallicity and the physical environment. The correction to the dust masses for CO contamination, and the revision to $H_2$ masses inferred from a downward revision of the X parameter (from 2.8 x $10^{20}$ $cm^{-2}$ (K $kms^{-1}$)$^{-1}$ to 2.7 x $10^{19}$ $cm^{-2}$ (K $kms^{-1}$)$^{-1}$ ) found by Yao et al. (2003), allow a re-examination of the g/d ratios derived from sub-mm methods for a sub-sample of SLUGS galaxies.

In general, significant uncertainties remain in determining the g/d ratio by any method because of errors in inferring both the gas and dust masses. It is generally thought that g/d is in the range 100-200 in our Galaxy (e.g. Hildebrand 1983, Devereux and Young 1990, Sodroski et al. 1997). Values for the local interstellar cloud are found in the range 100-190 (Frisch & Slavin 2002, Kimura et al. 2003). The values for starburst galaxies are much less clear, since there are particularly large uncertainties in estimating both gas and dust masses from mm, sub-mm and far-IR data. Values for g/d found are typically several hundred (see below). Using a different approach, with models of optical and IR data for the nuclear starburst in IR luminous galaxies, Contini & Contini (2003) find g/d highly variable from galaxy to galaxy and within galaxies with a total range 30 $\lesssim$ g/d $\lesssim$ 6000.

The dust and gas masses from Dunne et al. (2000) lead to g/d values averaging several hundred when the standard CO-to-$H_2$ conversion factor of X = 2.8x$10^{20}$ $cm^{-2}$ (K $kms^{-1})^{-1}$ is used (see section 1.2). The major source of uncertainty in the absolute dust masses from 850μm data is the value for the mass opacity coefficient $_{850}$. As noted by Dunne et al. (2000), observational estimates of $_{850}$ rely on an extrapolation to 850μm from 120-200μm using the dust emissivity index which is also uncertain. The uncertainty in $_{850}$ may be as high as a factor ~7 (Hughes, Dunlop & Rawlings (1997). The major source of uncertainty in gas masses is in the CO-to $H_2$ conversion factor X, especially for starburst galaxies. Accordingly, we first present our results, and then discuss their significance in the light of these uncertainties.

### 3.2.1 Ratio g/d measured globally

There are eighteen galaxies in the Yao et al. (2003) SLUGS sub-sample which have both integrated HI and CO(1-0) fluxes reported by Dunne et al. (2000). The CO fluxes were used by Dunne et al. to obtain global $H_2$ masses using the conventional X factor. The FIR luminosities for this sample lie in the range 1.2x$10^{10}$ - 8.7x$10^{11}$ $L_\odot$ (including Arp 220). From the global dust and gas masses reported in Dunne et al., the median g/d = 434 for this sub-sample based on their single temperature (isothermal) dust model, and g/d = 225 for their two component model incorporating a cold component at $T_d$ = 20K. All dust mass determinations assume = 2.0. We then applied factors of 0.75 and 0.62 respectively to the isothermal and two component dust masses (columns 4 and 5 of Table 4 in Dunne et al. (2000)) to correct for the CO contamination at the respective levels of 25 and 38 per cent (see section 3.1). Second, we applied the revised X factor (see section 1.2) to obtain revised molecular gas masses from the CO luminosities. This results in a downward revision of these masses by an order of magnitude. It is assumed here that these factors are applicable globally, and we obtain respective median values for g/d of 316 and 194 for the isothermal and two component dust models, with a dispersion of about a factor of three. Thus the effect of even a drastic reduction in X is not large, which is attributable to the large fraction of global gas mass in the form of HI, measured directly from 21cm line observations.

### 3.2.2 Ratio g/d in the central few kpc

A dramatically different picture emerges if we apply the same analysis to the nuclear regions. For this purpose we used a separate smaller sub-sample of twelve galaxies in the Yao et al. (2003) CO survey for which there are also HI VLA maps obtained by Thomas et al. (2002). The galaxies for this well resolved sample have a smaller range in FIR luminosity, corresponding to 1.2x$10^{10}$ - 6.2x$10^{10}$ $L_\odot$. The HI maps have resolution 25", comparable to the ~15" resolution of the 850μm data from Dunne et al. (2000) and the CO data from Yao et al. (2003). The range in projected beamsize corresponding to the JCMT measurements for this sub-sample is 2.5 – 4.7 kpc. The spatial resolution of the HI maps is sufficiently close to the JCMT beamsize to permit a reliable scaling the HI masses from 25" to 15" assuming a uniform brightness distribution. The $H_2$ masses within the ~15" beam of the JCMT may be derived from the CO luminosities $L_{CO(1\text{-}0)}$ in Table 2 of Yao et al. (2003) using the equation

$$M(H_2) = 2.3 \times 10^3 \left[ \frac{X}{2.7 \times 10^{19}} \right] L_{CO(1\text{-}0)} \; M_\odot \qquad (4)$$

The resulting masses are already reported by Yao et al., but Equation (4) is given to show how the masses scale with the assumed value for the parameter X. Equation (4) was derived using a gaussian beam solid angle $_b$ = 5.6x10$^{-9}$ sr corresponding to the FWHM beamsize of ~15". Corresponding dust masses for the ~15" beam were obtained by first scaling the isothermal global dust masses in column 4 of Table 4 in Dunne et al. (2000) by the ratio of the central brightness (corresponding to the measured CO position) to the integrated flux at 850µm using the SCUBA maps. Second, a factor of 0.75 was applied to correct for CO contamination appropriate for the isothermal model. We use only the isothermal dust model in this case since star formation is heavily concentrated in the nuclear regions, and it seems less likely that a cold component would coexist with the warm component.

The results for the nuclear sample are shown in Table 1, and yield a median value g/d = 50, with a dispersion of about a factor of two about the median. Without CO corrections to the dust mass and using the conventional X-factor, the median value g/d = 310. The more dramatic change after the corrections in this case arises because of the much larger fraction of gas in the form of $H_2$ in the nuclear region compared to that globally, giving the reduction in X a much greater impact.

There are only four galaxies common to both of the two sub-samples of eighteen and twelve - namely NGC 3221, 3583, 5600, and 5665. The corresponding average value for the global g/d is 420 before the corrections, and 360 after corrections for CO contamination and the lower X-factor. For the nuclear region g/d = 380 before correction, and 94 after the applying the above corrections.

Thus values of g/d derived for the well resolved sub-sample are close to 50, applicable to the inner few kpc. We re-emphasize here that the foregoing estimates for the nuclear regions are based on dust masses assuming a single temperature model (averaging ~ 35K) from Dunne et al. (2000). If the second colder dust component were included, it would raise the dust masses, and thus decrease the g/d ratios still further.

**3.2.3** Principal uncertainties and significance of the derived g/d ratios

To summarize, the values for the *global* g/d ratios derived from data reported in Dunne et al. (2000) are not significantly altered by corrections applied for CO line contamination and a reduction in the CO-to-$H_2$ conversion factor X. Median corrected values are in the range 200-300. In the *nuclear* region, however, the reduction is from ~ 300 to ~ 50 because of the larger fraction of gas in the form of $H_2$ and the large correction to the $H_2$ masses. The uncertainties in the absolute values of g/d may be large, however, and we address this issue next.

The uncertainty in   is unlikely to be large since DE01 have made an excellent case that  =2.0

for the SLUGS sample based on the consistency of the 850μm/450μm flux ratios. This result contributes to some reduction in the uncertainty in the extrapolation of the far-IR dust opacity to sub-mm wavelengths and to the determination of the dust temperature $T_d$, which is otherwise plagued by the degeneracy in determining and $T_d$ simultaneously. The dominant sources of uncertainty are the remaining error in $_{850}$ and in the CO-to-$H_2$ factor X. The uncertainty in the former is largely unknown, and, as noted earlier, has been estimated to be as high as a factor of 7. We are inclined to regard this as an extreme value in light of the current strong evidence that =2.0, which was used to extrapolate to 850μm from the far-IR. A factor of three seems more likely. The range in the factor X for the SLUGS CO sub-sample derived by Yao et al. (2003) is about a factor of 2, and we regard this as an appropriate value for its uncertainty. Combining these in quadrature (logarithmically), the resulting uncertainty in g/d corresponds to about a factor of ~ 3-4 depending on the fraction of gas in the form of $H_2$. It is important to note that the dust masses in Table 1 scale inverseley with $_{850}$ / 0.077 $m^2$ $kg^{-1}$ and the $H_2$ masses linearly with X / $2.7\times10^{19}$ $cm^{-2}$ (K-km $s^{-1}$)$^{-1}$, permitting the values of g/d to be adjusted for future revisions in these parameters.

The uncertainties notwithstanding, it is interesting that a value of g/d ~ 50 for the nuclear regions of these galaxies is lower than estimates for our own Galaxy, though meaningful comparisons would require an estimate for the central region of the Milky Way. The more significant result, however, is that the g/d ratios appear to be about a factor of 6 lower in the nuclear regions than in the outer parts of the SLUGS galaxies. This result is not sensitive to errors in the dust opacity, but remains sensitive (by up to a factor of 2) to the parameter X because of the different $H_2$/HI ratios in the two regions. There are two possible explanations for this result. On the one hand, it would be consistent with higher metallicity in the central regions of star forming galaxies. An equally possible explanation is that considerable amounts of cold dust remain hidden in the outer non-starburst regions of these galaxies, and that the global g/d ratios are accordingly overestimated. The latter reason would be consistent with recent results for disk galaxies based on spatially resolved studies of their sub-mm SEDs (DE01 and references therein). Only a sensitive search for cold dust in the outer regions will settle the latter issue.

Finally, we return briefly to the results of Contini & Contini (2003) who find a much larger range of g/d within the nuclear regions of galaxies, based on model fitting to their SED's, primarily in the MIR, and to optical emission line ratios. Comparisons with our results are unfortunately difficult or impossible because their results are based upon much higher spatial resolution (< 1"), and appear to reflect regional variations on a much finer scale. Secondly, the ISM in these galaxies is effectively transparent at mm and sub-mm wavelengths whereas the results of Contini & Contini (2003) are applicable only to comparatively un-obscured regions. We conclude that no meaningful comparisons can be made in this instance.

## 4 CONCLUSIONS

Based on a survey of CO reported in Yao et al. (2003), we conclude that the average contribution by CO(3-2) emission to the SCUBA 850μm fluxes in SLUGS sample of IR galaxies measured by Dunne et al. (2000) is about 25 per cent. If this waveband lies predominantly in the Rayleigh-Jeans regime of the dust emission, then this factor is representative of the correction to the inferred dust masses. If a cold component with $T_d$ = 20K is included, then the reduction of 25 per cent in flux translates to an average of 38 percent in the dust mass. The shape of the dust mass functions presented in Dunne et al. (2000) and DE(01) are unaffected in by these results because the fractional correction appears to be uncorrelated with the dust mass. However, the Schechter parameter M* should be reduced by 25 per cent and 38 per cent for the isothermal and two component dust models respectively. We also estimate from the excitation ratios CO(3-2)/CO(1-0) in Yao et al. (2003) that it is unlikely that any correction is required for the contribution by CO(6-5) to 450μm fluxes of the SLUGS galaxies measured with SCUBA.

The gas-to-dust mass ratios g/d for a sub-sample of SLUGS galaxies were derived using corrections to the 850μm fluxes for CO line contamination together with a newly derived mean value for the CO-to-$H_2$ conversion factor X for the SLUGS sample reported by Yao et al. (2003). The latter parameter is a factor of 10 lower than the conventional value used for giant molecular clouds in the Milky Way. Neither of these corrections has any major bearing on the *global* g/d ratios. The global ratios are several hundred regardless of whether the corrections are applied or not since most of the global gas is HI. However, based on a limited sample of 12 galaxies with higher resolution HI data, the median value of g/d in the *central few kpc* is much lower (~50) than the global values (~300), which is attributable to the low value derived for X combined with the larger fraction of gas in molecular form. While we regard a figure of g/d near 50 to be plausible for the nuclear regions, the absolute uncertainties are large, and more significance should be attached to the variation between the nuclear and outer regions. Such a strong variation within galaxies could be introduced by either a strong metallicity gradient or by large amounts of undetected cold dust in the outer regions of these galaxies.

## ACKNOWLEDGEMENTS

ERS would like to thank Dr. David Naylor, University of Lethbridge, for providing access to his measurements of the SCUBA 850μm bandpass filter, and Dr. Slavek Rucinski, University of Toronto, for assistance with some aspects of the data reduction. LY thanks the Kavli Institute for Theoretical Physics, University of California at Santa Barbara, for their generous hospitality during the preparation of this paper. We thank the referee for useful suggestions which led to a significant clarification of the paper. This research was performed with the support of a grant from the Natural Sciences and Engineering Research Council of Canada.

Table 1. Dust and gas masses

| Name | Projected beam (kpc) (1) | $M(H_2)$ ($10^8$ $M_\odot$) (2) | M(HI) ($10^8$ $M_\odot$) (3) | $M_d$ ($10^6$ $M_\odot$) (4) | g/d (5) |
|---|---|---|---|---|---|
| NGC 3110 | 4.8 | 5.1 | 1.9 | 23 | 30 |
| NGC 3221 | 3.9 | 2.3 | 8.5 | 7.0 | 154 |
| NGC 3583 | 2.1 | 0.8 | 0.3 | 2.7 | 41 |
| NGC 3994/5 | 3.1 | 1.8 | 1.9 | 7.5 | 49 |
| NGC 5104 | 5.3 | 7.8 | 1.5 | 12 | 77 |
| UGC 8739 | 4.8 | 4.5 | 1.4 | 23 | 26 |
| NGC 5433 | 4.2 | 1.8 | 2.6 | 11 | 40 |
| NGC 5600 | 2.2 | 0.2 | 0.7 | 1.4 | 64 |
| NGC 5665 | 2.1 | 1.5 | 0.6 | 1.8 | 117 |
| NGC 5900 | 2.4 | 1.9 | 0.1 | 4.1 | 49 |
| NGC 5936 | 3.8 | 3.6 | 1.0 | 9.0 | 51 |
| NGC 6052 | 4.5 | 2.7 | 6.0 | 8.3 | 104 |

(1) Projected beam diameter based on FWHM 15".
(2) Mass of $H_2$ derived from Yao et al. (2003).
(3) Mass of HI within 15" beam scaled from maps presented by Thomas et al. (2002).
(4) Dust mass from col (4) of Dunne et al. (2000) corrected for CO and scaled to 15".
(5) Gas/dust mass ratio.

# FIGURE CAPTIONS

## Figure 1

The profile of the SCUBA filter (dashed) used in the SLUGS survey by Dunne et al. (2000), expressed as a function of recession velocity, with the ordinate in arbitrary units. The filter data have been smoothed to a velocity resolution of 300 km $s^{-1}$. The smooth curve (solid) is a least squares fit of a double gaussian to the data, used to derive the function F(V) expressed in equation (2).

## Figure 2

Plot of SCUBA equivalent flux S(V) produced by the CO(3-2) line vs. SCUBA flux at the central pixel for sources in the CO survey by Yao et al. (2003). Filled circles correspond to $V < 10{,}000$ km $s^{-1}$ and unfilled circles to $V > 10{,}000$ km $s^{-1}$. The straight line through the origin is a least squares fit to the data with $V < 10{,}000$ km $s^{-1}$ and has a slope of $0.26 \pm 0.01$. Some points were excluded from the fit. See text for details.

## Figure 3

The distribution of the ratios of SCUBA equivalent flux produced by the CO(3-2) line to the central pixel SCUBA flux for the Yao et al. (2003). These data are restricted to those objects with SCUBA brightness with $S/N > 3$ and $V < 10{,}000$ km $s^{-1}$. The dashed curve through the data represents the expected distribution for a true ratio of $r = 0.24$ and mean fractional uncertainties of respectively 32 per cent and 25 per cent for the SCUBA equivalent fluxes S(V) (numerator) and the single pixel SCUBA fluxes (denominator). See text for details.

## Figure 4

Plot of the ratio of CO(3-2) integrated brightness $I_{32}$ vs. SCUBA central pixel flux for all data with central pixel SCUBA brightness having $S/N > 3$ , regardless of recession velocity.

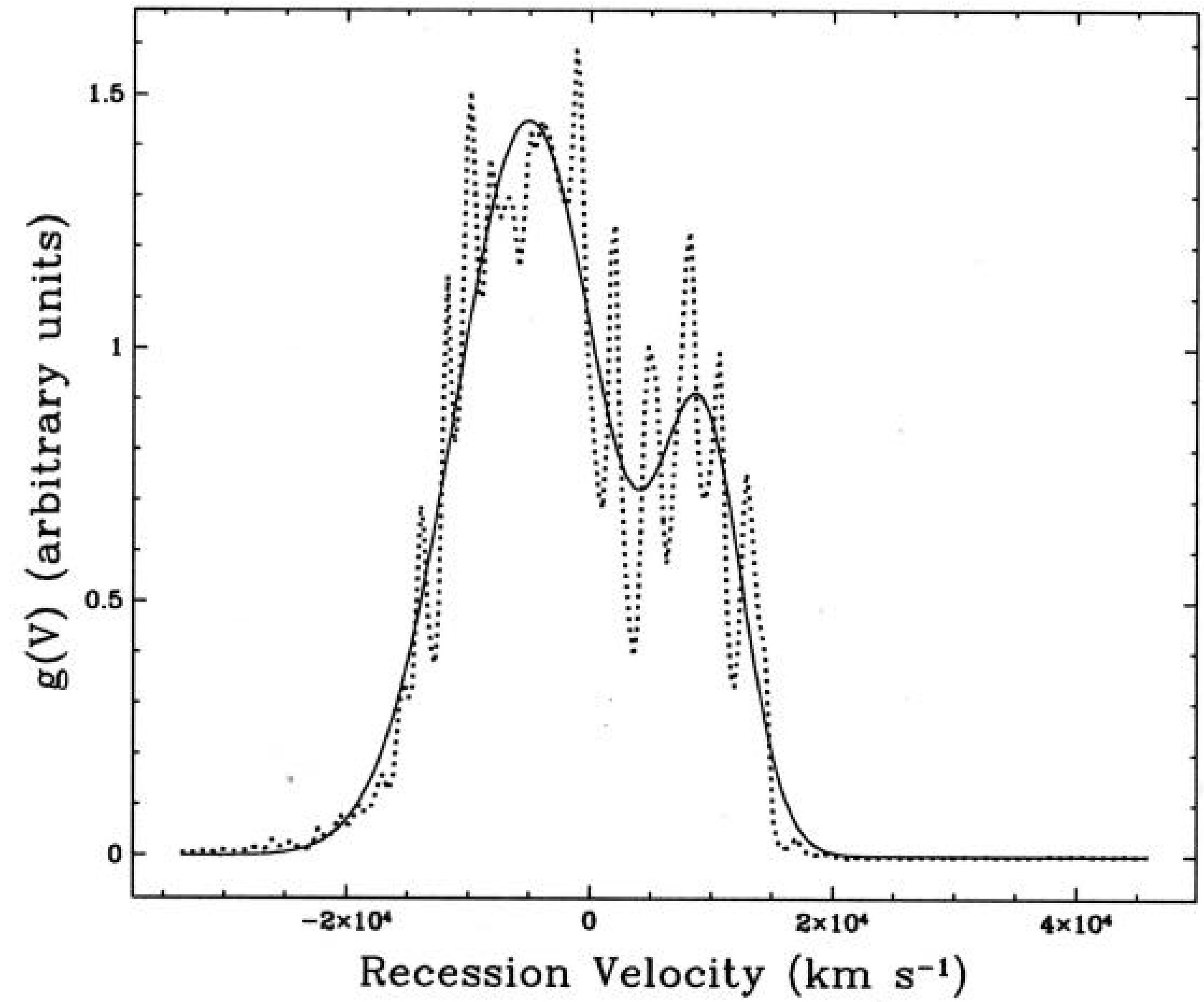
1.5
1
0.5
0
g(V) (arbitrary units)
−2×10⁴
0
2×10⁴
4×10⁴
Recession Velocity (km s⁻¹)

Figure 1

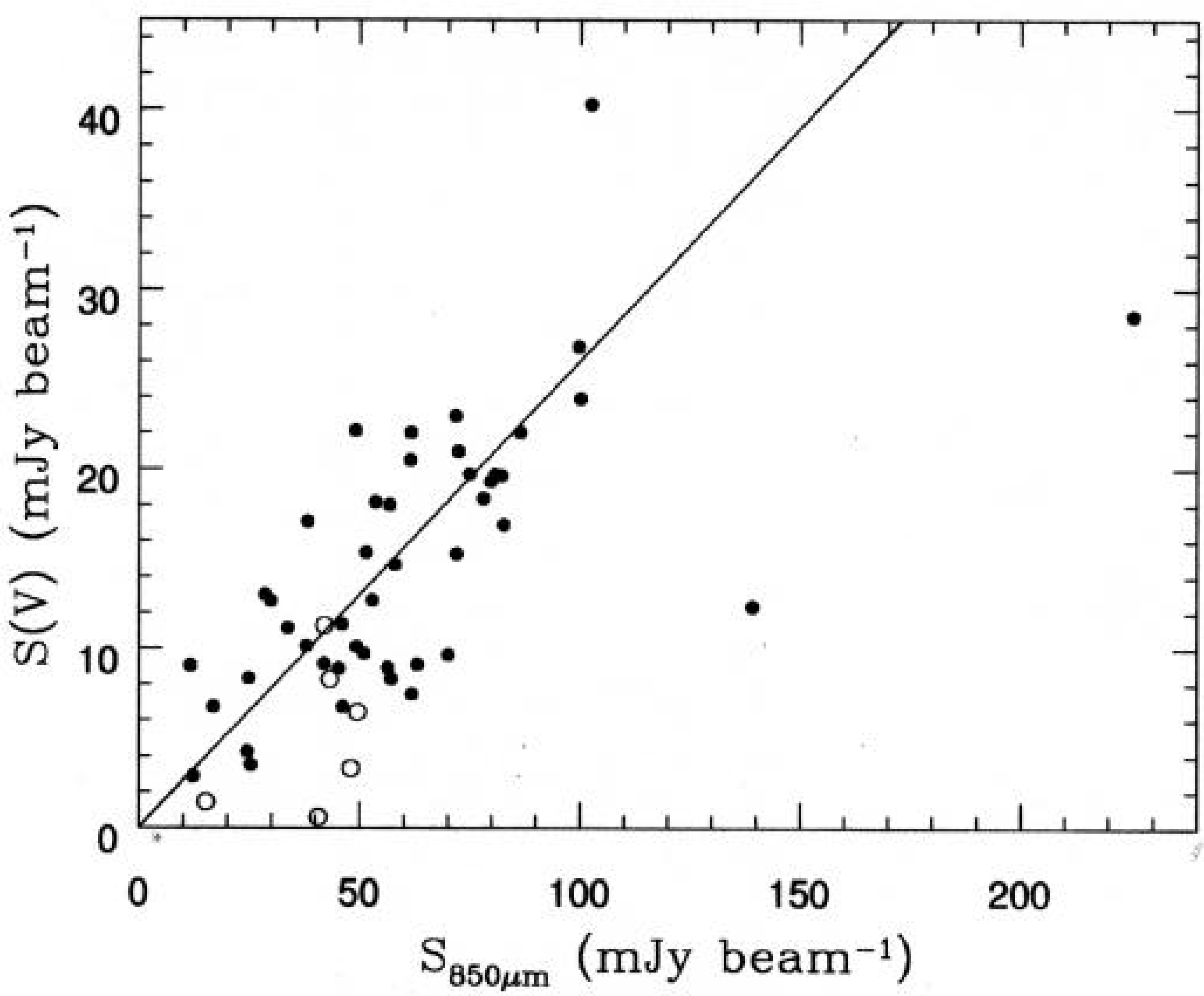

Figure 2

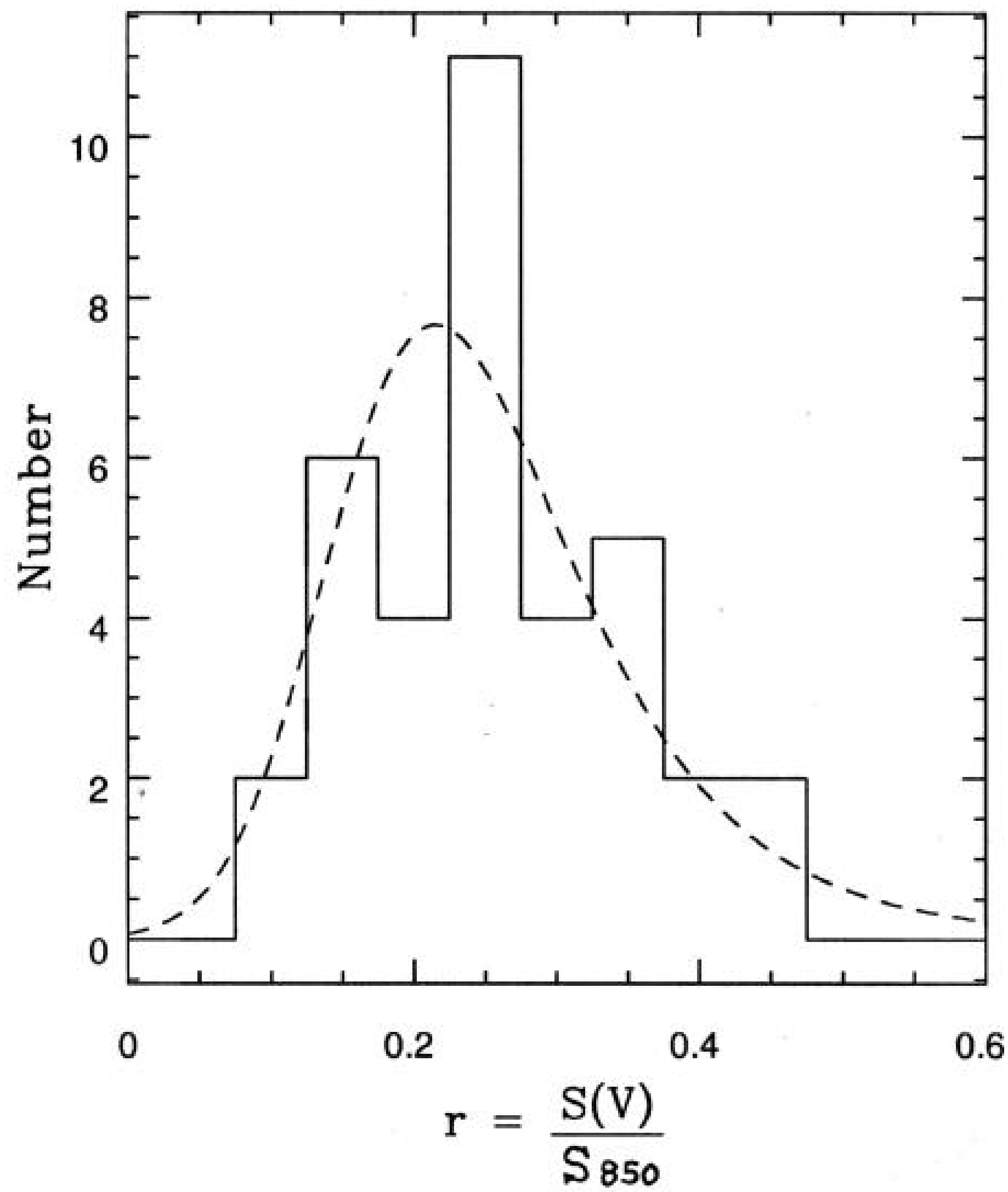

Figure 3

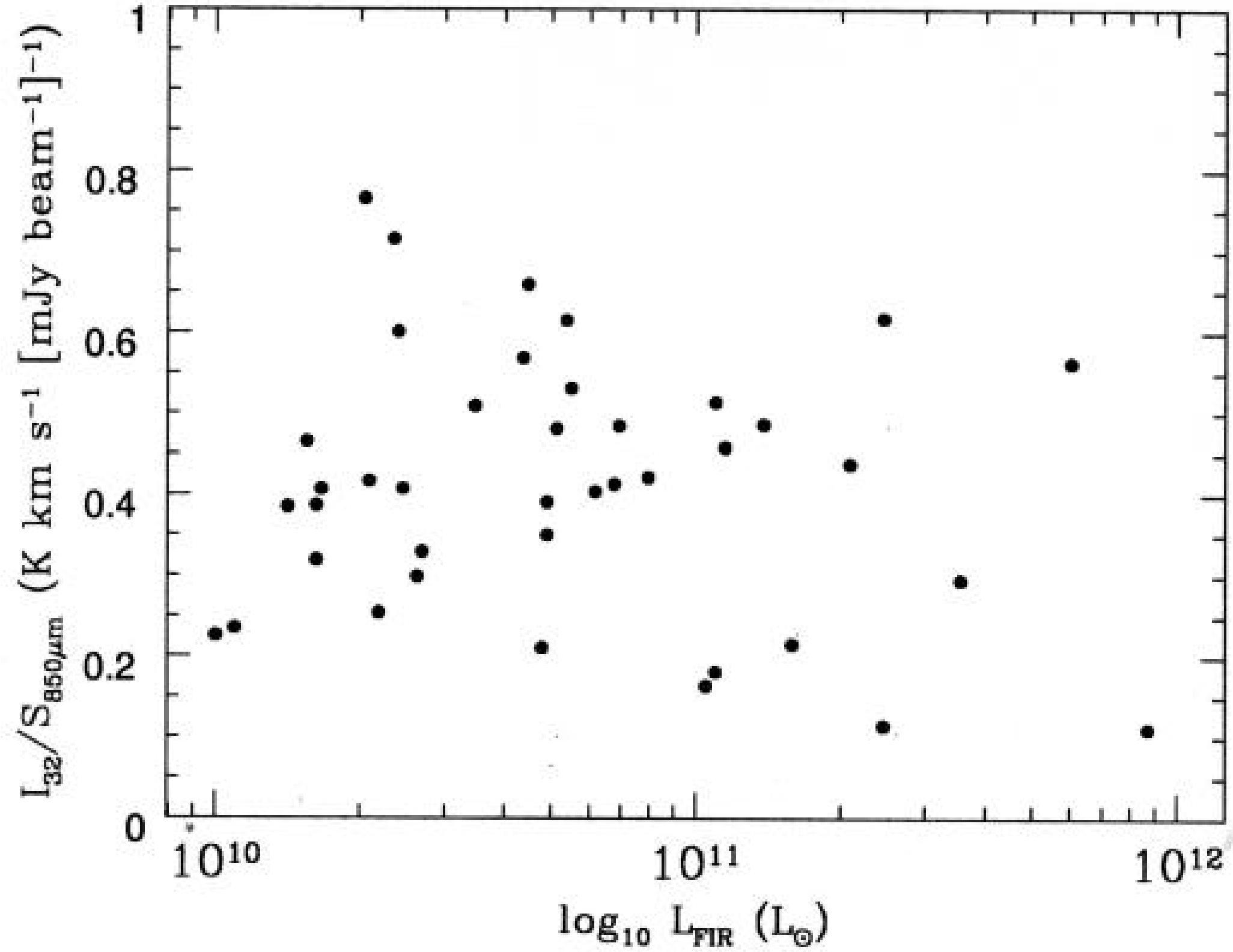
$I_{32}/S_{850\mu m}$ (K km s$^{-1}$ [mJy beam$^{-1}$]$^{-1}$)
0
0.2
0.4
0.6
0.8
1
$10^{10}$
$10^{11}$
$10^{12}$
$\log_{10} L_{FIR}$ ($L_\odot$)

Figure 4